\documentclass[english]{article}
\usepackage{mathptmx}

\usepackage[T1]{fontenc}
\usepackage[utf8]{luainputenc}
\usepackage[a4paper]{geometry}
\usepackage{babel}
\usepackage{array}
\usepackage{float}
\usepackage{multirow}
\usepackage{graphicx}
\usepackage[unicode=true,pdfusetitle,
 bookmarks=true,bookmarksnumbered=false,bookmarksopen=false,
 breaklinks=false,pdfborder={0 0 1},backref=false,colorlinks=false]
 {hyperref}

\makeatletter

\providecommand{\tabularnewline}{\\}

\usepackage{babel}

\makeatother

\begin{document}

\title{Probing small $x(g)$ region with the LHeC based $\gamma p$ colliders}

\author{U. Kaya$^{1}$, S. Sultansoy$^{1,2}$, G. Unel$^{3}$\\
\textit{\small $^{1}$TOBB University of Economics and Technology,
Ankara, Turkey}\\
\textit{\small $^{2}$ANAS Institute of Physics, Baku, Azerbaijan}\\
\textit{\small $^{3}$UC Irvine,USA}}

\date{$\,$}
\maketitle
\begin{abstract}
Understanding of the parton distributions at small $x(g)$ is one
of the most important issues for clarfiying of the QCD basics. In
this paper potential of the LHeC for probing small $x(g)$ region
via $c\bar{c}$ and $b\bar{b}$ production has been investigated.
Comparison of the $ep$ and real $\gamma p$ options of the LHeC clearly
show the advantage of $\gamma p$ collider option. Measurement of
$x(g)$ down to $3\times10^{-6}$ with high statistics, especially
at $\gamma p$ option, seems to be reachable which is two order smaller
than HERA coverage.
\end{abstract}

\section{Introduction}

The problem of precise measurement of parton distribution functions
(PDF) is yet to be solved for the energy scales relevant to the LHC
results. On the other hand, precison knowledge on the parton distribution
of small $x_{B}$ and sufficiently large $Q^{2}$ is crucial for enlightening
the QCD basics at all levels, from partons to nuclei. Besides, with
the recent discovery of the 125 GeV scalar particle \cite{ATLAS Higgs},
\cite{CMS Higgs} at the LHC, the basic components of the electroweak
part of the Standard Model (SM) have been completed. Hovewer, the
Higgs mechanism provides less than 2\% of mass of the visible universe.
Remaining 98\% are provided by the QCD part of the SM. Therefore,
clarifying of the basics of the QCD is important for a better understanding
of our universe. That's why the QCD explorer was proposed ten years
ago (see review \cite{QCD 2004} and references therein). One of the
required measurements is the gluon PDF for low momentum fraction:
small $x(g)$. The last machine that has probed $x(g)$ was HERA which
had a reach of about $x(g)>$$10{}^{-4}$. Large Hadron-Electron Collider
(LHeC) project \cite{J. Phys. G: Nucl. Part. Phys. 39 (2012) 075001}
- the most powerful microscope ever designed - will provide a unique
opportunity to probe extremely small $x(g)$ region. In this project,
where proton-electron collisions are aimed the $e-beam$ can be obtained
from a new circular or linear machine.

Today, LR option is considered as the basic one for the LHeC \cite{arXiv:1211.4831v1 [hep-ex] 20 Nov 2012}.
Actually this decision was almost obvious from the beginning due to
the complications in constructing by-pass tunnels around the existing
experimental caverns and installing the $e-ring$ in the already commissioned
tunnel. Let us remind that the CDR stage of the LHC assumed also $ep$
collisions using the already existed LEP ring; but it turned out that
LHC installation required dismantling of LEP from the tunnel.

Within the linac-ring option of the LHeC, a proton beam from LHC can
be hit with a high energy electron or photon beam. The photons may
be virtual ones from the electron beam resulting in a typical DIS
event or these can be real photons originating from the Compton Back
Scattering process. In the latter case, the photon spectrum consists
of the high energy photons peaking at about 80\% of the electron beam
energy on the continuum of Weizsacker-Williams photons. The present
study aims to investigate the feasibility of a small $x(g)$ measurement
with such a machine. Main parameters of $ep$ and $\gamma p$ options
of the LHeC are presented in section 2. Section 3 is devoted to investigation
of small $x(g)$ region using the processes $\gamma p$ $\rightarrow$
$c\bar{{c}}X$ and $\gamma p$ $\rightarrow$ $b\bar{{b}}X$. The
generator level results are obtained using CompHEP \cite{CompHep}
software package. Comparison with processes $ep$ $\rightarrow$ $ec\bar{{c}}X$
and $ep$ $\rightarrow$ $eb\bar{{b}}X$ shows an obvious advantage
of the LHeC $\gamma p$ option, which will provide more than one order
higher cross sections at small $x(g)$ region comparing to the $ep$
option. Finally, section 4, provides a summary of the conclusion together
with some suggestions.

\section{Main parameters of $ep$ and $\gamma p$ options of the LHeC}

It should be emphasized that real $\gamma p$ collisions can be achieved
only on the base of linac ring type $ep$ colliders (see review \cite{A. N. Akay}
for history and status of linac-ring type collider proposals). Real
$\gamma$ beam for $\gamma p$ collider \cite{S. I. Alekhin}, \cite{A. K. Ciftci et al.},
\cite{TESLA * HERA based}, \cite{Conversion efficiency} will be
produced using the Compton back scattering of laser beam off the high
energy electron beam \cite{I. F. Ginzburg}, \cite{Principles of photon colliders}.
Possible application of this mechanism to the other LHeC option under
consideration, namely to ring-ring type $ep$ colliders results in
negligible $\gamma p$ luminosities, $L_{\gamma p}$$<10{}^{-7}$$L{}_{ep}$.

Currently, two versions for the $ep$ option of the LHeC are under
consideration: multi-pass energy recovery linac (ERL) yielding $L_{ep}=$$10^{33}$$cm^{-2}$$s{}^{-1}$
and pulsed single pass linac yielding $L{}_{ep}=$$10{}^{32}$$cm{}^{-2}$$s{}^{-1}$.
In the first case, $E{}_{e}=60\, GeV$ has been chosen as a base electron
energy, since higher energies are not available because of the synchrotron
radiation loss in the arcs. In the second case, beam energies above
$140\, GeV$ would be available \cite{J. Phys. G: Nucl. Part. Phys. 39 (2012) 075001}.
These two options will be denoted as LHeC-1 and LHeC-2. Main parameters
of the LHeC $ep$ collisions, in different options are presented in
Table 1.

\begin{table}[H]
\begin{centering}
\caption{Main parameters of ep collisions.}

\par\end{centering}

\begin{centering}
\begin{tabular}{|c|c|c|c|c|}
\hline 
 & $E_{e},\, GeV$ & $E_{p},\, TeV$ & $\sqrt{s},\, TeV$ & $L,\, cm^{-2}$$s{}^{-1}$\tabularnewline
\hline 
\hline 
ERL & $60$ & $7$ & $1.30$ & $10^{33}$\tabularnewline
\hline 
LHeC-1 & $60$ & $7$ & $1.30$ & $9\times$$10^{31}$\tabularnewline
\hline 
LHeC-2 & $140$ & $7$ & $1.98$ & $4\times10^{31}$\tabularnewline
\hline 
\end{tabular}
\par\end{centering}

\label{TABLE1}
\end{table}

In the $\gamma p$ option the luminosity of $\gamma p$ collisions
will be similar to the luminosity of $ep$ collisions for the pulsed
single straight linac. In the ERL case, L$_{\gamma p}$ will be 10
times lower than L$_{ep}$ as the energy recovery does not work after
Compton back scattering.

\section{Inclusive processes yielding $c\bar{c}$ and $b\bar{b}$ final states
at LHeC}

\begin{figure}[H]
\begin{centering}
\includegraphics[width=0.4\paperwidth]{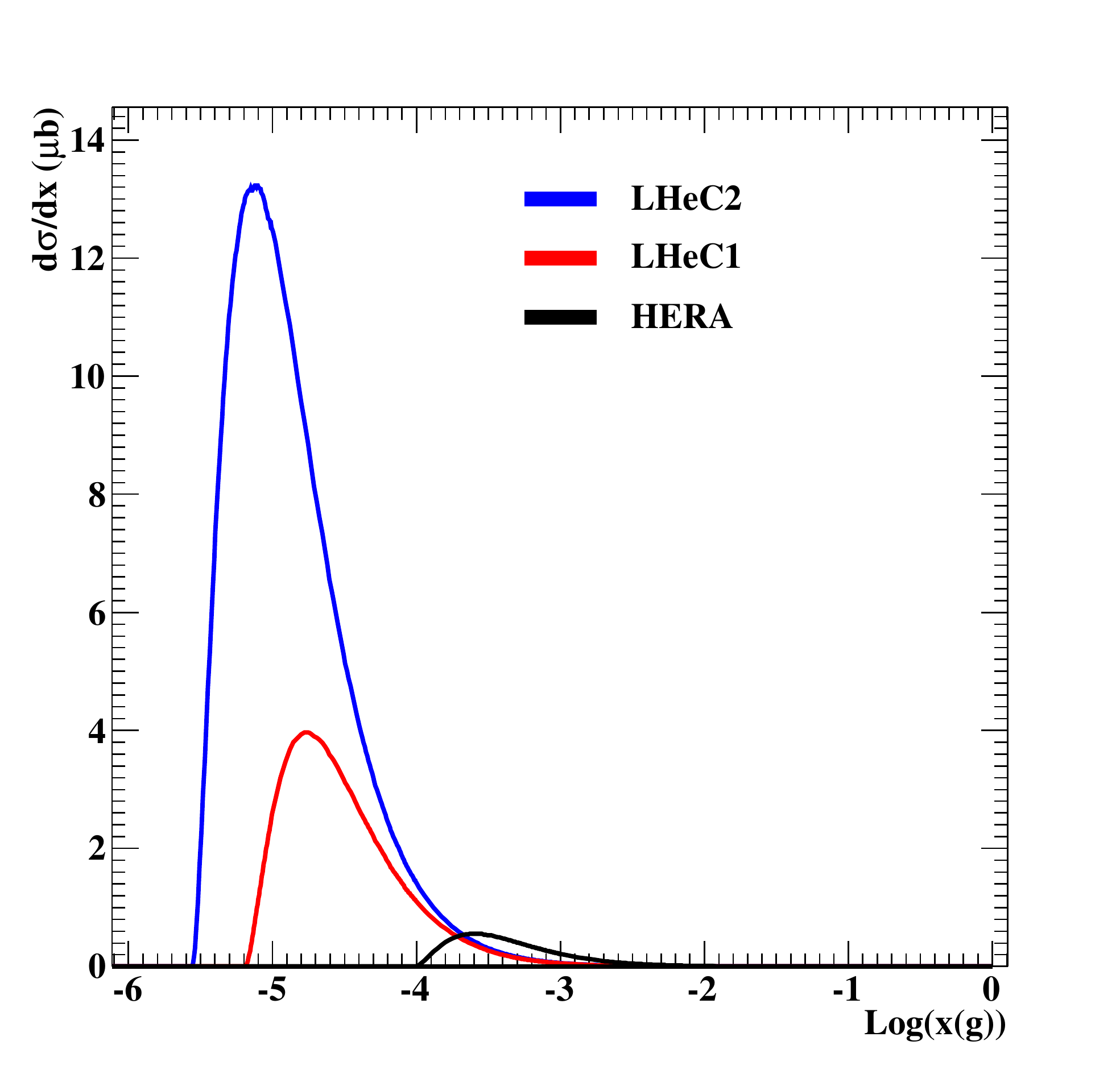}\includegraphics[width=0.4\paperwidth]{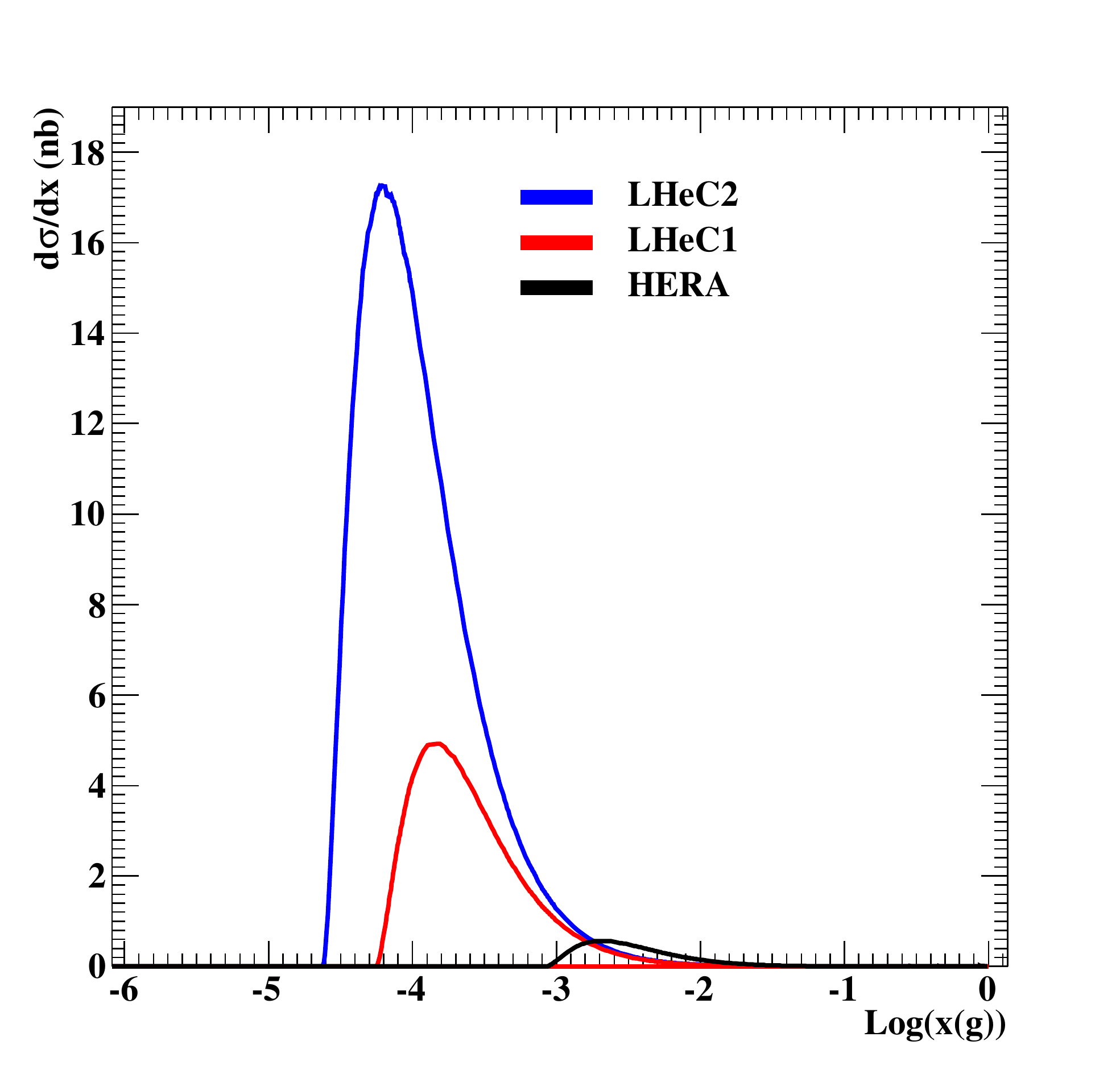}
\par\end{centering}

\caption{The x(g) reach and differential cross sections in $c\bar{{c}}$ (left)
and $b\bar{{b}}$ (right) final states for the HERA and the LHeC. }

\label{FIGURE1} 
\end{figure}

The final states that can be easily distinguished from the background
events and that would give a good measure of the $x(g)$ are $eg\to eq\bar{q}$
and/or $\gamma g\to q\bar{q}$ where the gluon ($g$) is from the
LHC protons, electrons and photons are from a new accelerator (namely,
an electron linac providing beams tangential to the LHC) to be build
and the letter $q$ stands for a heavy quark flavour, such as $b$
quark and possibly $c$ as well. The $b$ quark final states are easier
to identify due to $b$-tagging possibility using currently available
technologies: for example, ATLAS silicon detectors have about 70\%
$b$-tagging efficiency. In Table \ref{TABLE2} we present the cross
sections for heavy quark pair production via DIS, quasi real photons
(WWA) and Compton Back Scattering (CBS) photons at the LHeC with $E{}_{e}=60\, GeV$
and $E{}_{e}=140\, GeV$. For comparison, we also give values for
DIS and WWA processes at HERA. It is seen that WWA quasi real photons
are advantageous comparing to DIS and CBS photons are advantageous
comparing to WWA. All numerical calculations are performed using CompHep
\cite{CompHep} with CTEQ6L1 \cite{CTEQ} PDF distributions. In Figure
\ref{FIGURE1}, the differential cross section depending on the $x(g)$
has been shown for WWA photons at HERA and at LHeC. As expected, LHeC
will give opportunity to investigate an order smaller $x(g)$ than
HERA. 

\begin{table}[H]
\caption{Heavy quark pair production cross sections via DIS, WWA, and CBS mechanisms.}

\begin{centering}
\begin{tabular}[b]{|c|c|c|c|c|c|c|}
\hline 
\multirow{2}{*}{} & \multicolumn{3}{c|}{} & \multicolumn{3}{c|}{}\tabularnewline
 & \multicolumn{3}{c|}{$b\bar{b}\,(pb)$} & \multicolumn{3}{c|}{$c\bar{c}\,(pb)$}\tabularnewline
\hline 
Machine & DIS & WWA & CBS & DIS & WWA & CBS\tabularnewline
\hline 
HERA & $6.07\times10^{2}$ & $4.57\times10^{3}$ & - & $4.66\times10^{4}$ & $7.29\times10^{5}$ & -\tabularnewline
\hline 
LHeC-1($E{}_{e}=60$$\, GeV$) & $4.26\times10^{3}$ & $2.99\times10^{4}$ & $2.41\times10^{5}$ & $2.38\times10^{5}$ & $3.44\times10^{6}$ & $2.38\times10^{7}$\tabularnewline
\hline 
LHeC-2($E{}_{e}=140\, GeV$) & $7.07\times10^{3}$ & $4.91\times10^{4}$ & $3.70\times10^{5}$ & $3.72\times10^{5}$ & $5.27\times10^{6}$ & $3.46\times10^{7}$\tabularnewline
\hline 
\end{tabular}
\par\end{centering}

\label{TABLE2}
\end{table}

\begin{figure}[H]
\begin{centering}
\includegraphics[width=0.4\paperwidth]{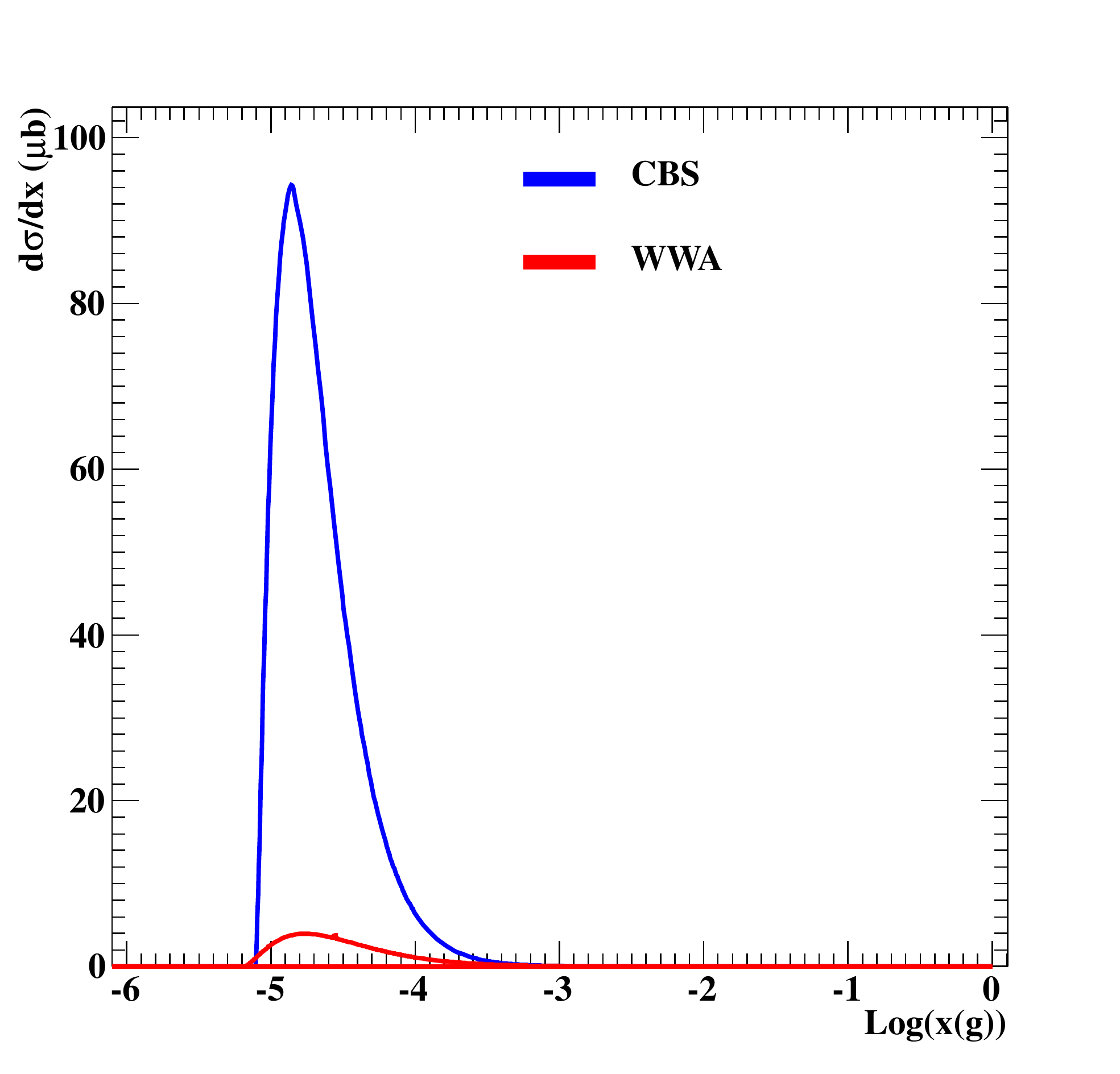}\includegraphics[width=0.4\paperwidth]{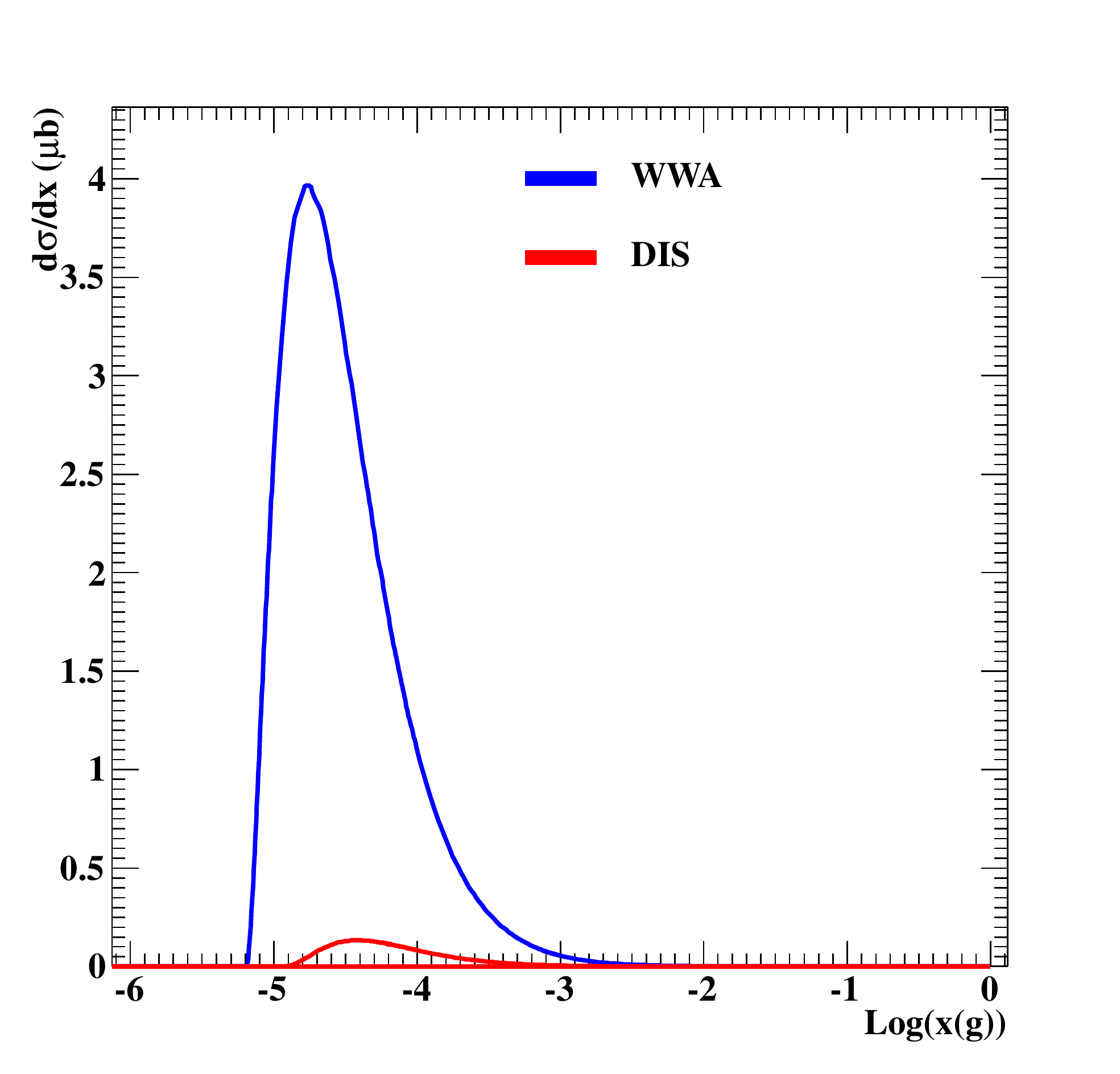} 
\par\end{centering}

\caption{Differential cross sections for $c\bar{{c}}$ final states produced
via CBS, WWA and DIS at the LHeC-1.}

\label{FIGURE2} 
\end{figure}

\begin{figure}[H]
\begin{centering}
\includegraphics[width=0.4\paperwidth]{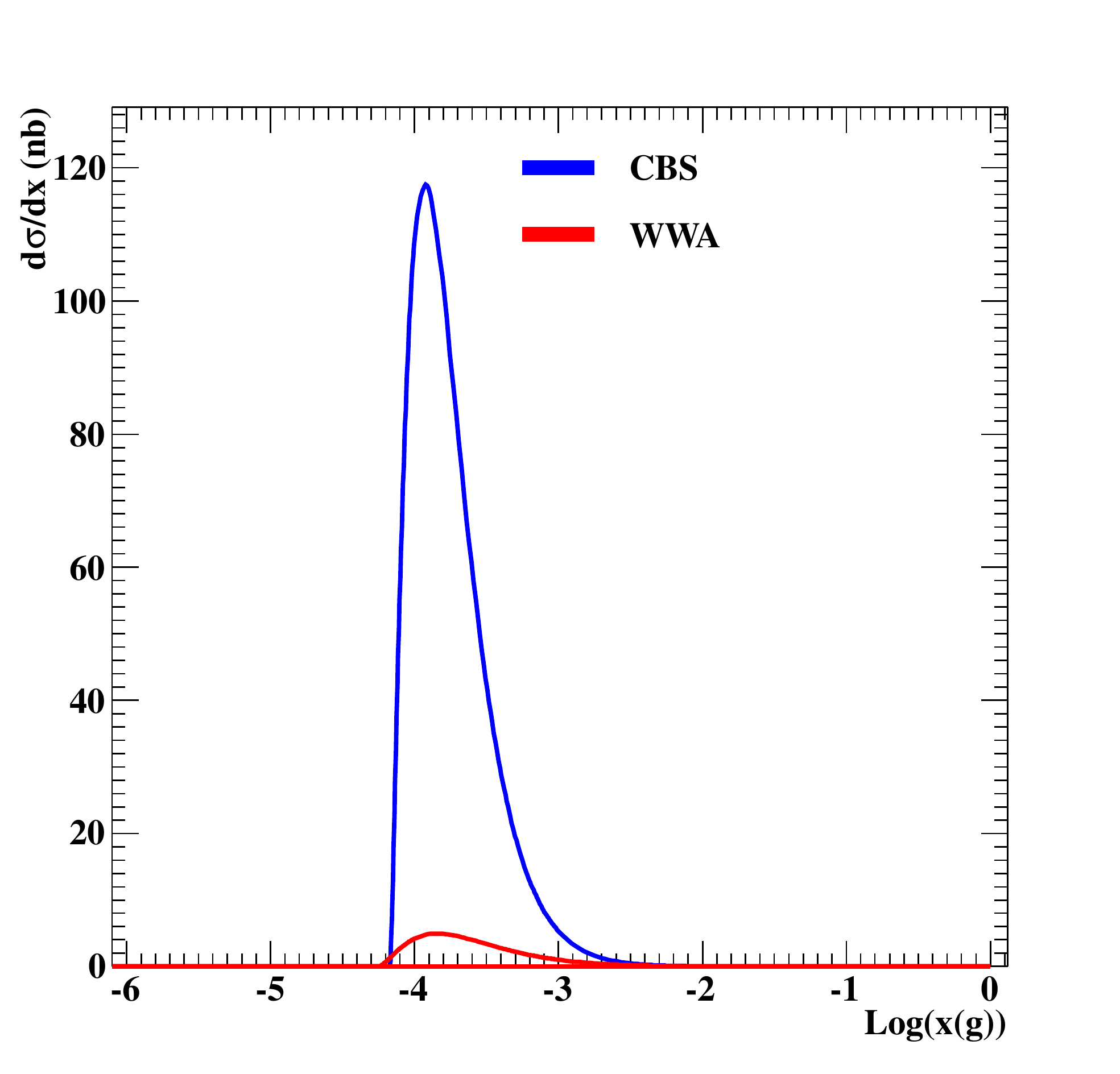}\includegraphics[width=0.4\paperwidth]{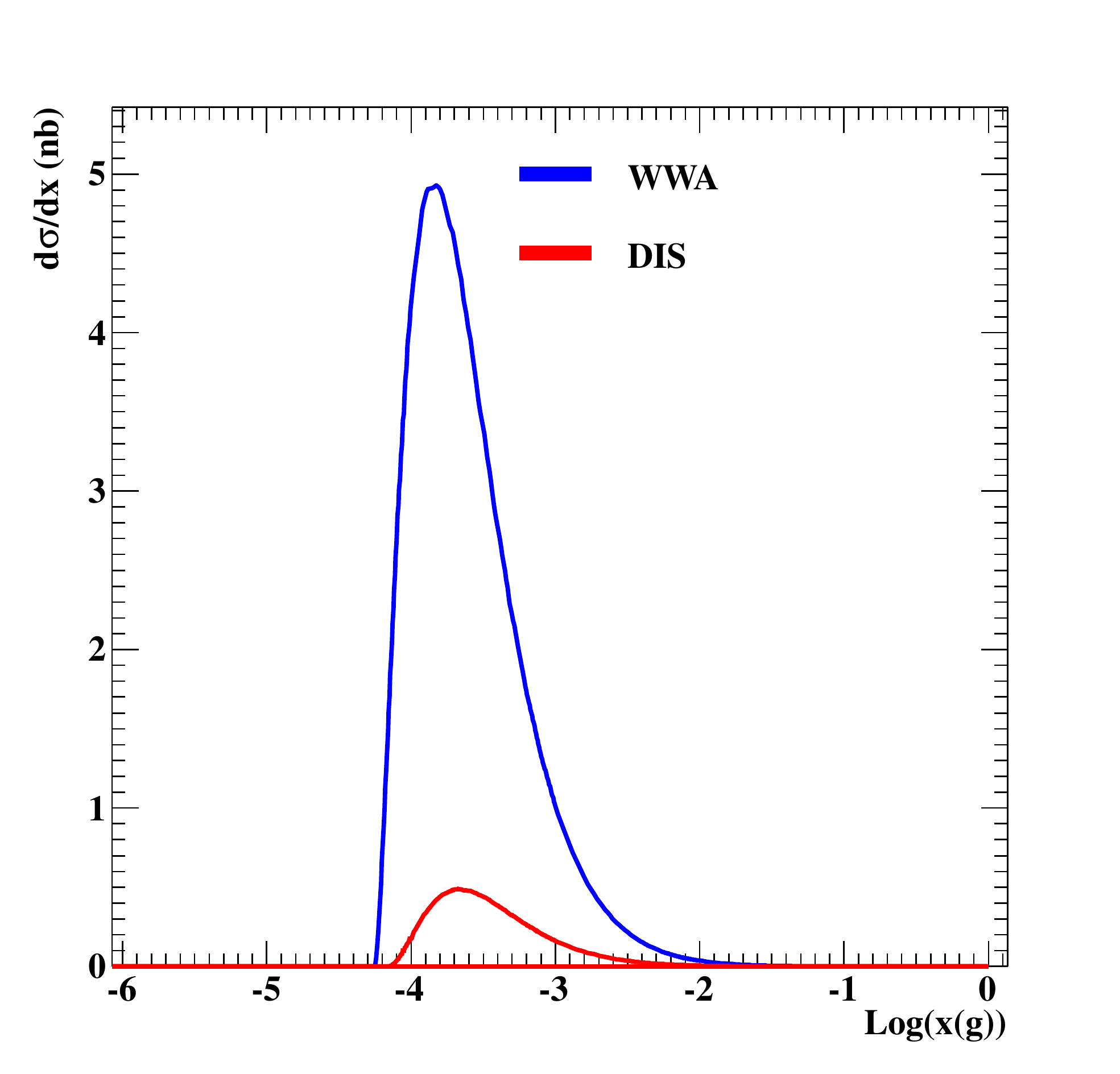}
\par\end{centering}

\caption{Differential cross sections for $b\bar{{b}}$ final states produced
via CBS, WWA and DIS at the LHeC-1.}

\label{FIGURE3}
\end{figure}

\begin{figure}[H]
\begin{centering}
\includegraphics[width=0.4\paperwidth]{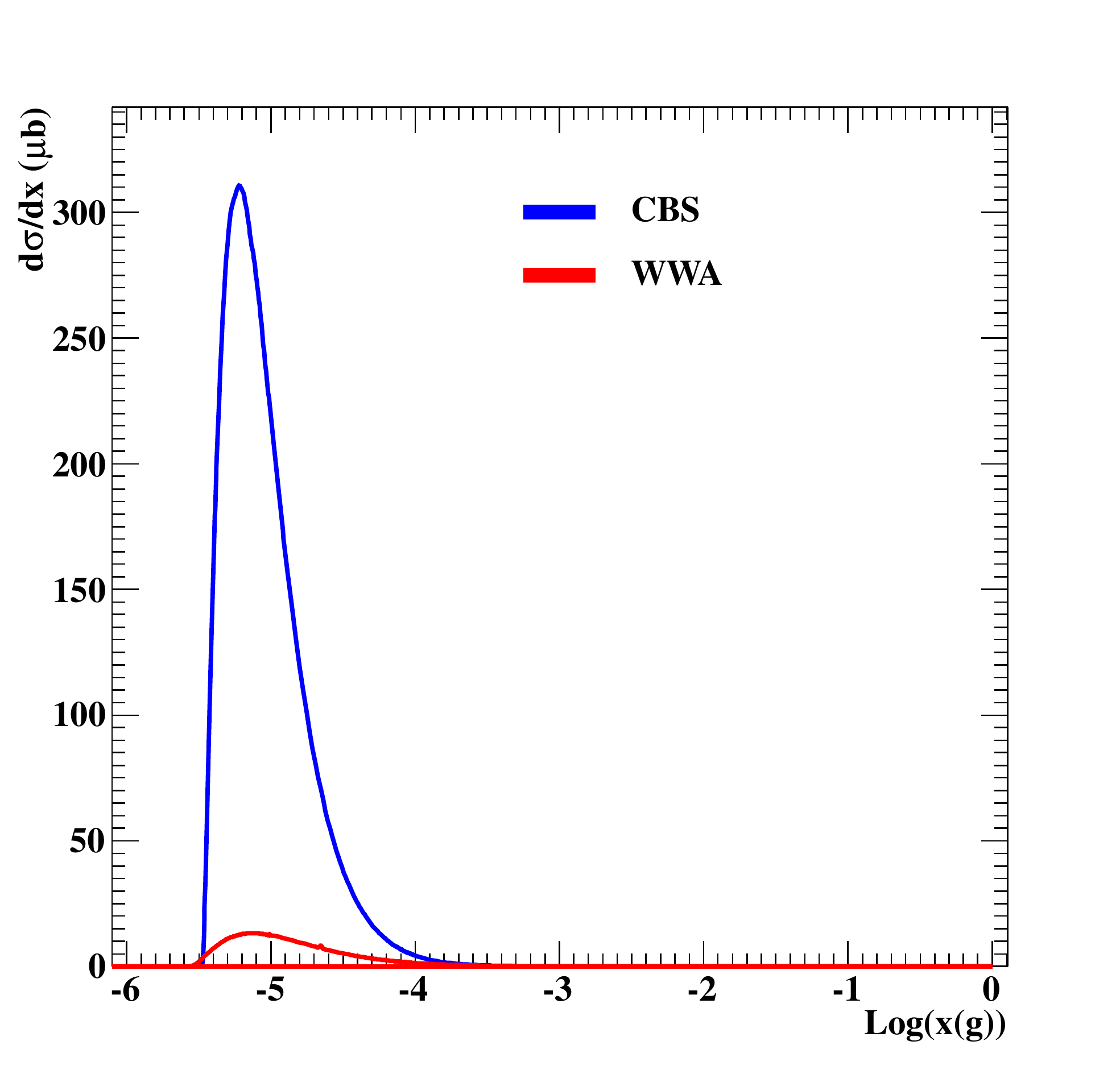}\includegraphics[width=0.4\paperwidth]{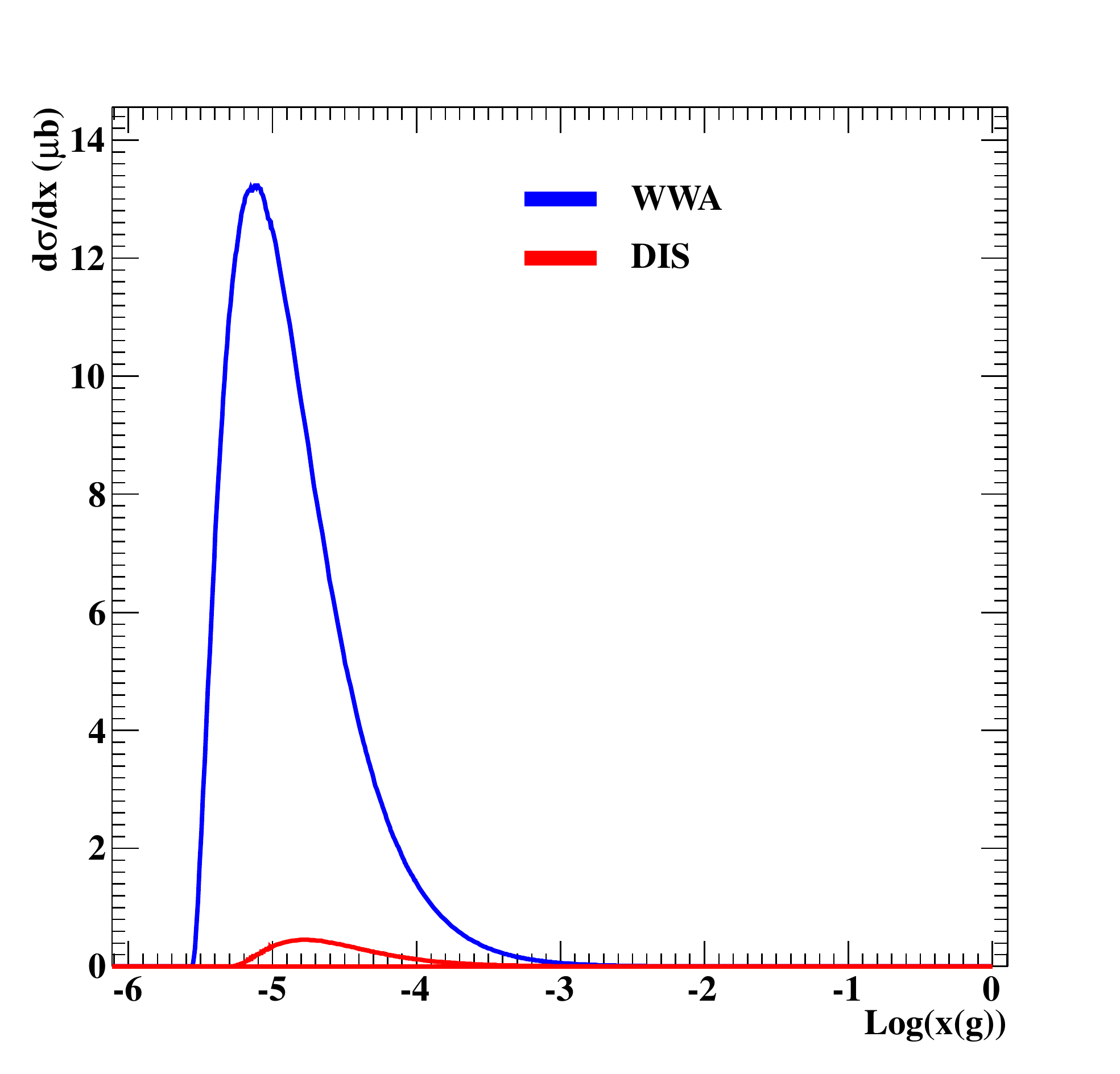} 
\par\end{centering}

\caption{Differential cross sections for $c\bar{{c}}$ final states produced
via CBS, WWA and DIS at the LHeC-2.}

\label{FIGURE4} 
\end{figure}

\begin{figure}[H]
\begin{centering}
\includegraphics[width=0.4\paperwidth]{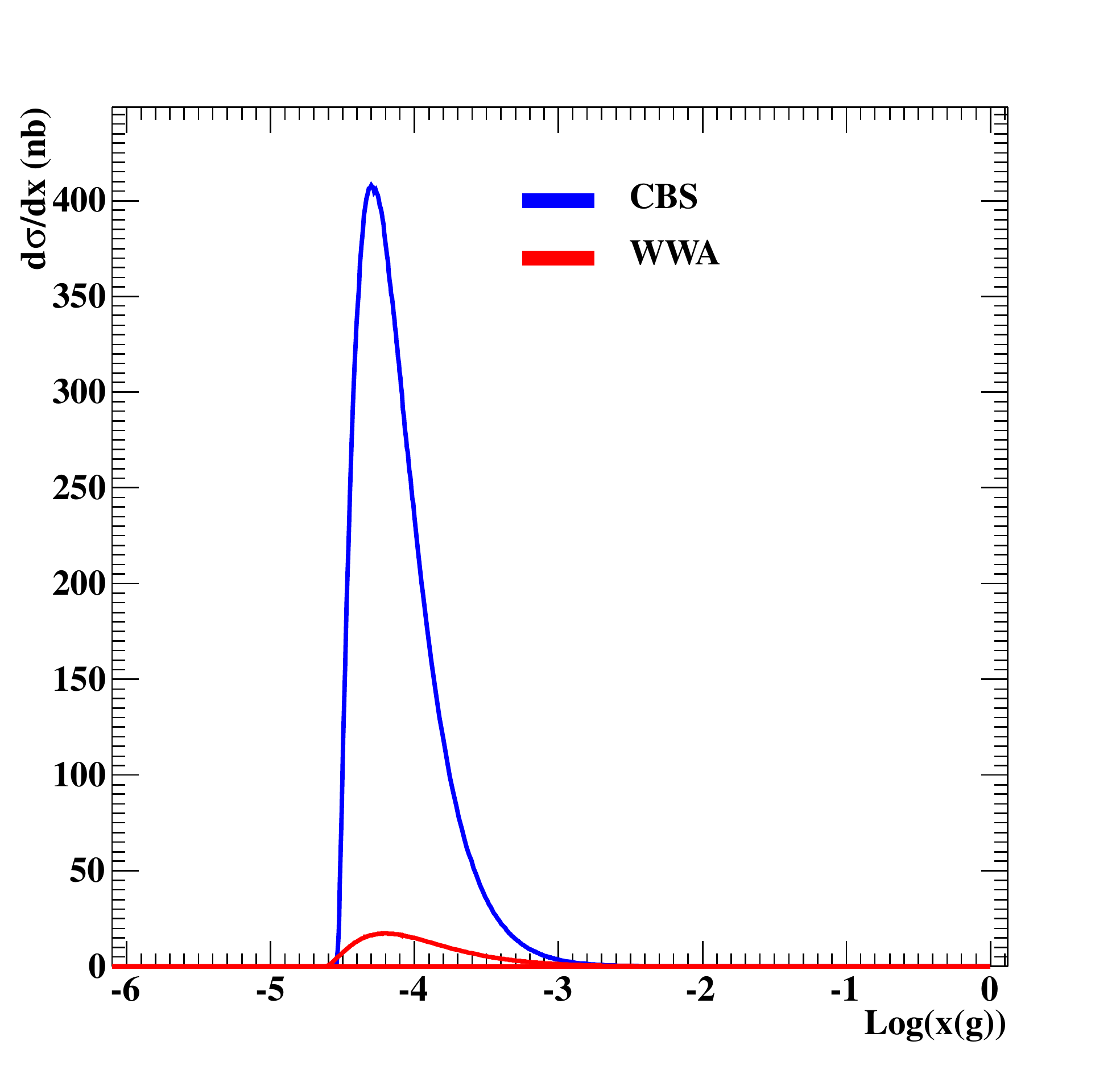}\includegraphics[width=0.4\paperwidth]{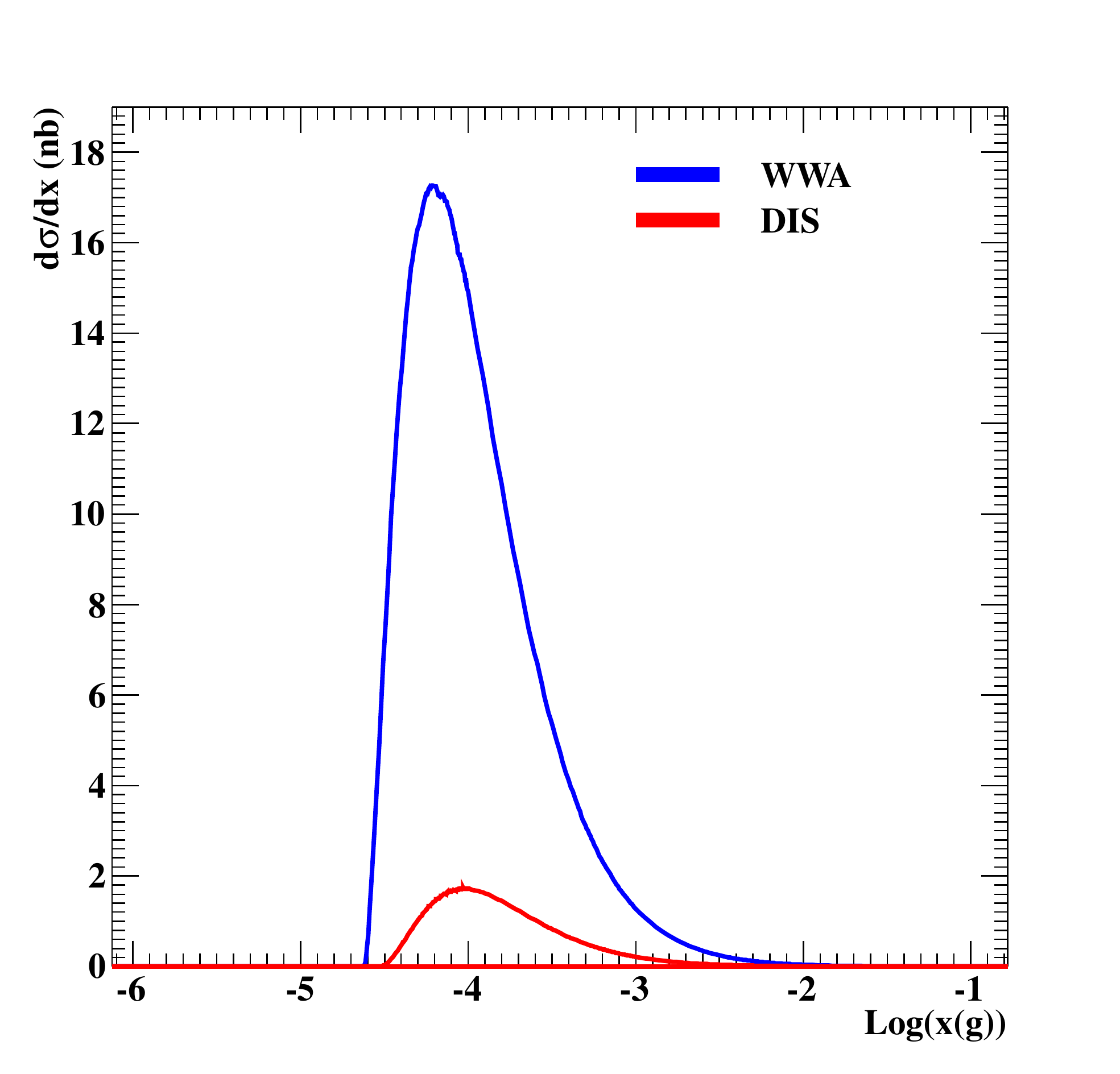}
\par\end{centering}

\caption{Differential cross sections for $b\bar{{b}}$ final states produced
via CBS, WWA and DIS at the LHeC-2 .}

\label{FIGURE5}
\end{figure}

The advantage of the CBS photons becomes even more obvious if one
analyzes $x(g)$ distribution of differential cross sections for CBS,
WWA and DIS. In Figure \ref{FIGURE2}, we show the $d\sigma/$$dx(g)$
at the LHeC-1 for $c\bar{c}$ production. It is seen that CBS at small
$x(g)$ region provides more than one (two) order higher cross sections
comparing to WWA (DIS). For example, differential cross section of
$c\bar{c}$ pair production at the LHeC-1 achieves maximum value $94\,\mu b$
at $x(g)=1.44\times10^{-5}$ for CBS, whereas maximum value for WWA
and DIS are $4\,\mu b$ at $x(g)=$$1.54\times10^{-5}$and $0.15\,\mu b$
at $x(g)=$$3.89\times10^{-5}$, respectively. Similar distributions
for $b\bar{b}$ at LHeC-1, $c\bar{c}$ at LHeC 2 and $b\bar{b}$ at
LHeC-2 are shown in Figures \ref{FIGURE3}, \ref{FIGURE4} and \ref{FIGURE5},
respectively. Maximum values of differential cross sections and corresponding
$x(g)$ values for DIS, WWA, and CBS at the LHeC-1 (2) are given in
the Table \ref{TABLE 3} (\ref{TABLE 4}). The advantage of CBS due
to large cross section is obvious from the comparison.

\begin{table}[H]
\caption{Maximum values of differential cross sections and corresponding $x(g)$
values for DIS, WWA, and CBS at the LHeC-1.}

\begin{centering}
\begin{tabular}{|c|c|c|c|c|}
\hline 
\multirow{2}{*}{} & \multicolumn{2}{c|}{} & \multicolumn{2}{c|}{}\tabularnewline
 & \multicolumn{2}{c|}{$c\bar{c}$ } & \multicolumn{2}{c|}{$b\bar{b}$}\tabularnewline
\hline 
 & $d\sigma/$$dx$ & $x$ & $d\sigma/$$dx$ & $x$\tabularnewline
\hline 
DIS & $0.15\,\mu b$ & $3.89\times10^{-5}$ & $0.47\, nb$ & $1.99\times10^{-4}$\tabularnewline
\hline 
WWA & $4.0\,\mu b$ & $1.54\times10^{-5}$ & $5.02\, nb$ & $1.25\times10^{-4}$\tabularnewline
\hline 
CBS & $94\,\mu b$ & $1.44\times10^{-5}$ & $117\, nb$ & $1.23\times10^{-4}$\tabularnewline
\hline 
\end{tabular}
\par\end{centering}

\label{TABLE 3}
\end{table}

\begin{table}[H]
\caption{Maximum values of differential cross sections and corresponding $x(g)$
values for DIS, WWA, and CBS at the LHeC-2.}

\begin{centering}
\begin{tabular}{|c|c|c|c|c|}
\hline 
\multirow{2}{*}{} & \multicolumn{2}{c|}{} & \multicolumn{2}{c|}{}\tabularnewline
 & \multicolumn{2}{c|}{$c\bar{c}$ } & \multicolumn{2}{c|}{$b\bar{b}$}\tabularnewline
\hline 
 & $d\sigma/$$dx$ & $x$ & $d\sigma/$$dx$ & $x$\tabularnewline
\hline 
DIS & $0.44\,\mu b$ & $1.54\times10^{-5}$ & $1.73\, nb$ & $9.12\times10^{-5}$\tabularnewline
\hline 
WWA & $13.2\,\mu b$ & $6.45\times10^{-6}$ & $17\, nb$ & $5.88\times10^{-5}$\tabularnewline
\hline 
CBS & $312\,\mu b$ & $6.02\times10^{-6}$ & $408\, nb$ & $5.01\times10^{-5}$\tabularnewline
\hline 
\end{tabular}
\par\end{centering}

\label{TABLE 4}
\end{table}

\begin{figure}[H]
\begin{centering}
\includegraphics[width=0.4\paperwidth]{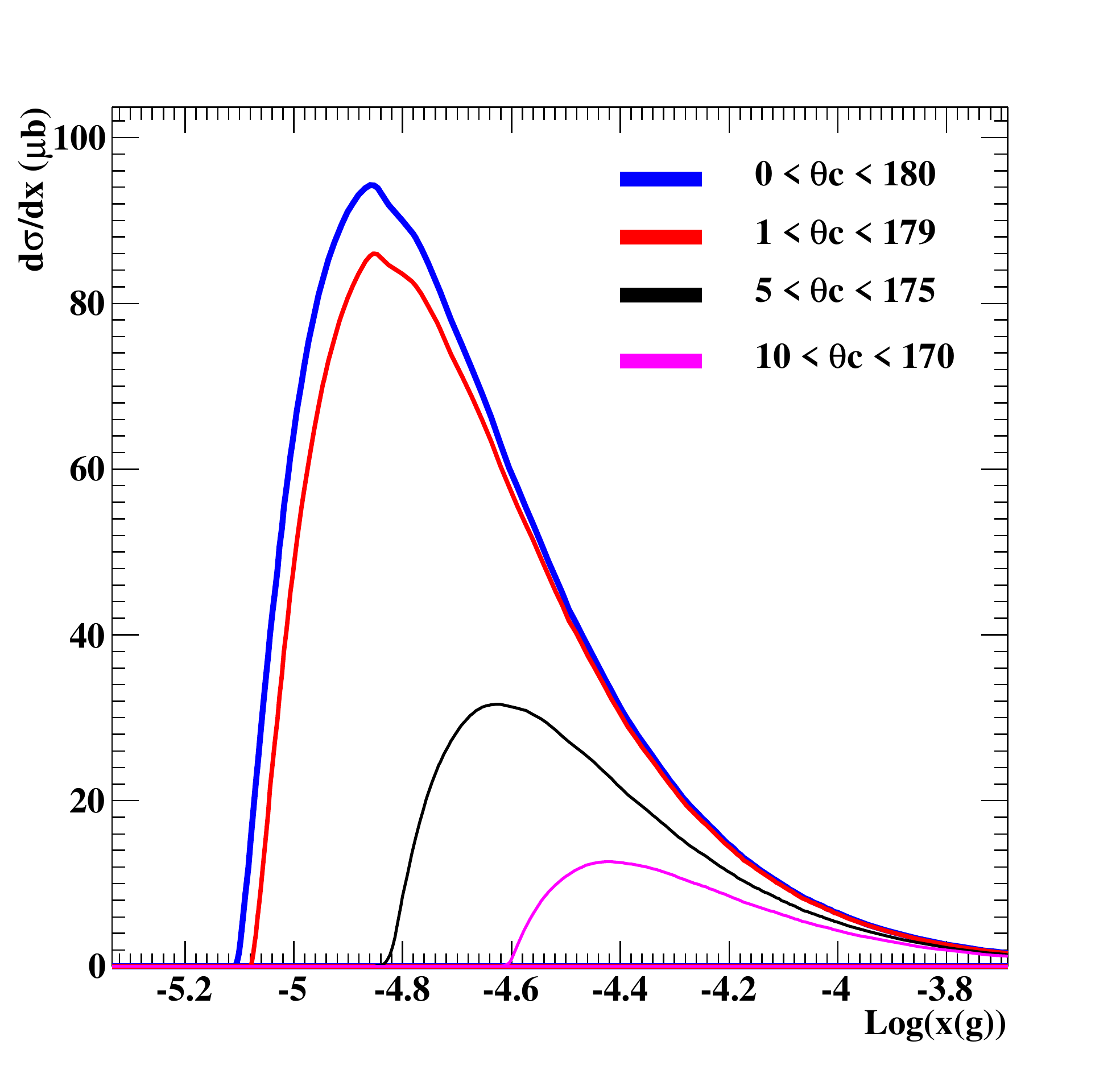}\includegraphics[width=0.4\paperwidth]{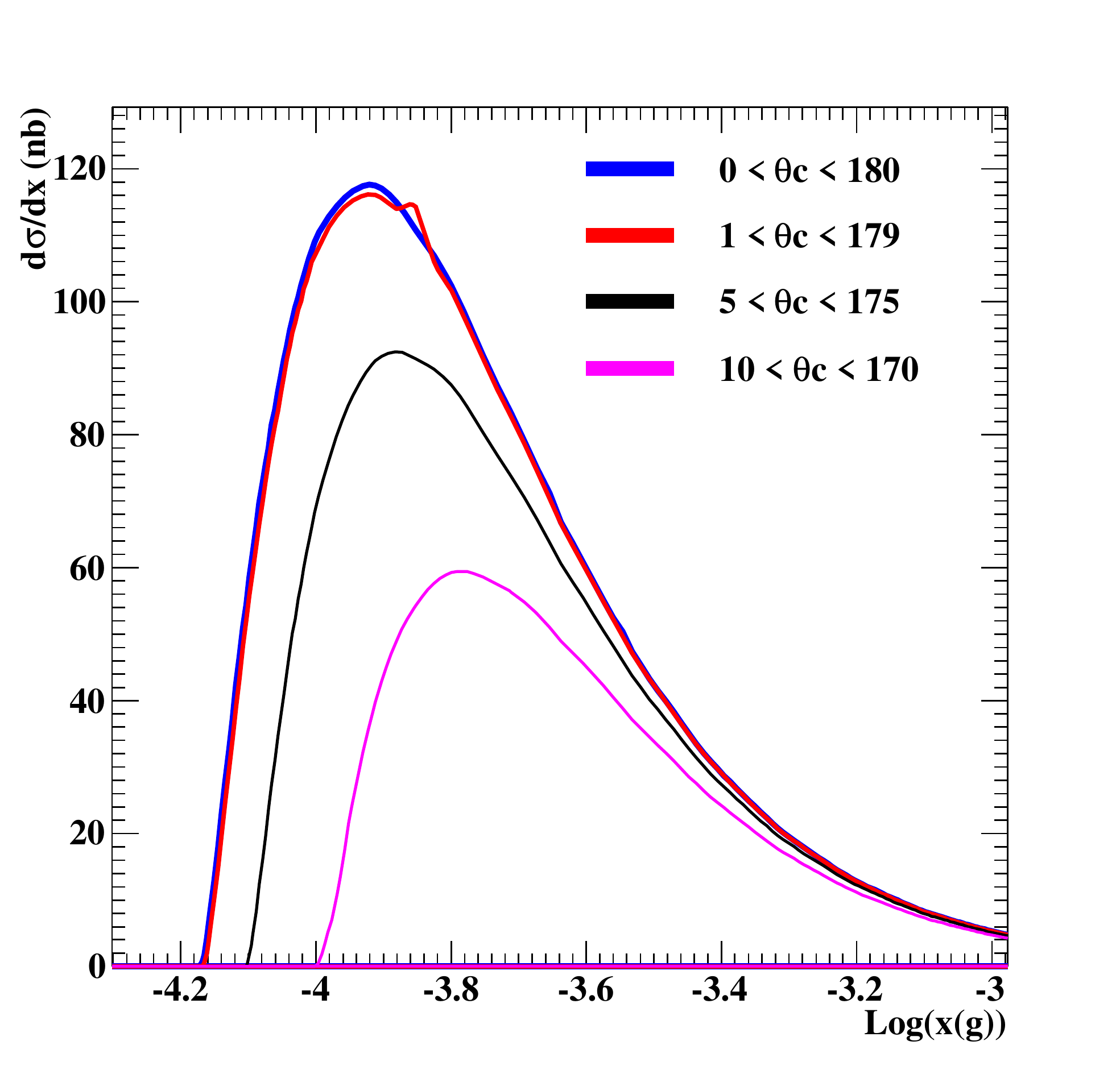}
\par\end{centering}

\caption{The effect of angular reach for $c\bar{c}$ (left) and $b\bar{b}$
(right) final states produced via CBS at the LHeC-1.}

\label{FIGURE6}
\end{figure}

\begin{figure}[H]
\begin{centering}
\includegraphics[width=0.4\paperwidth]{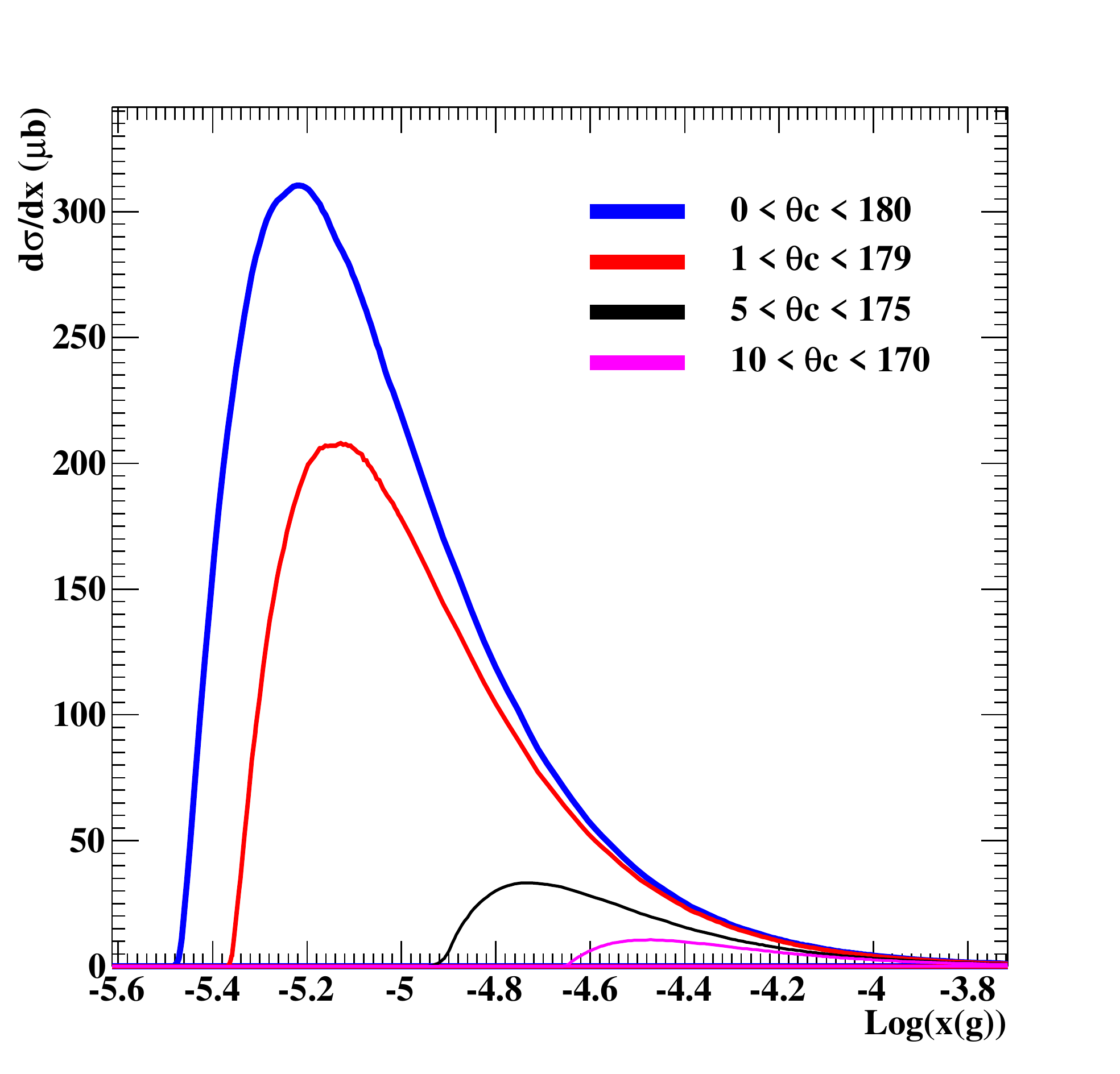}\includegraphics[width=0.4\paperwidth]{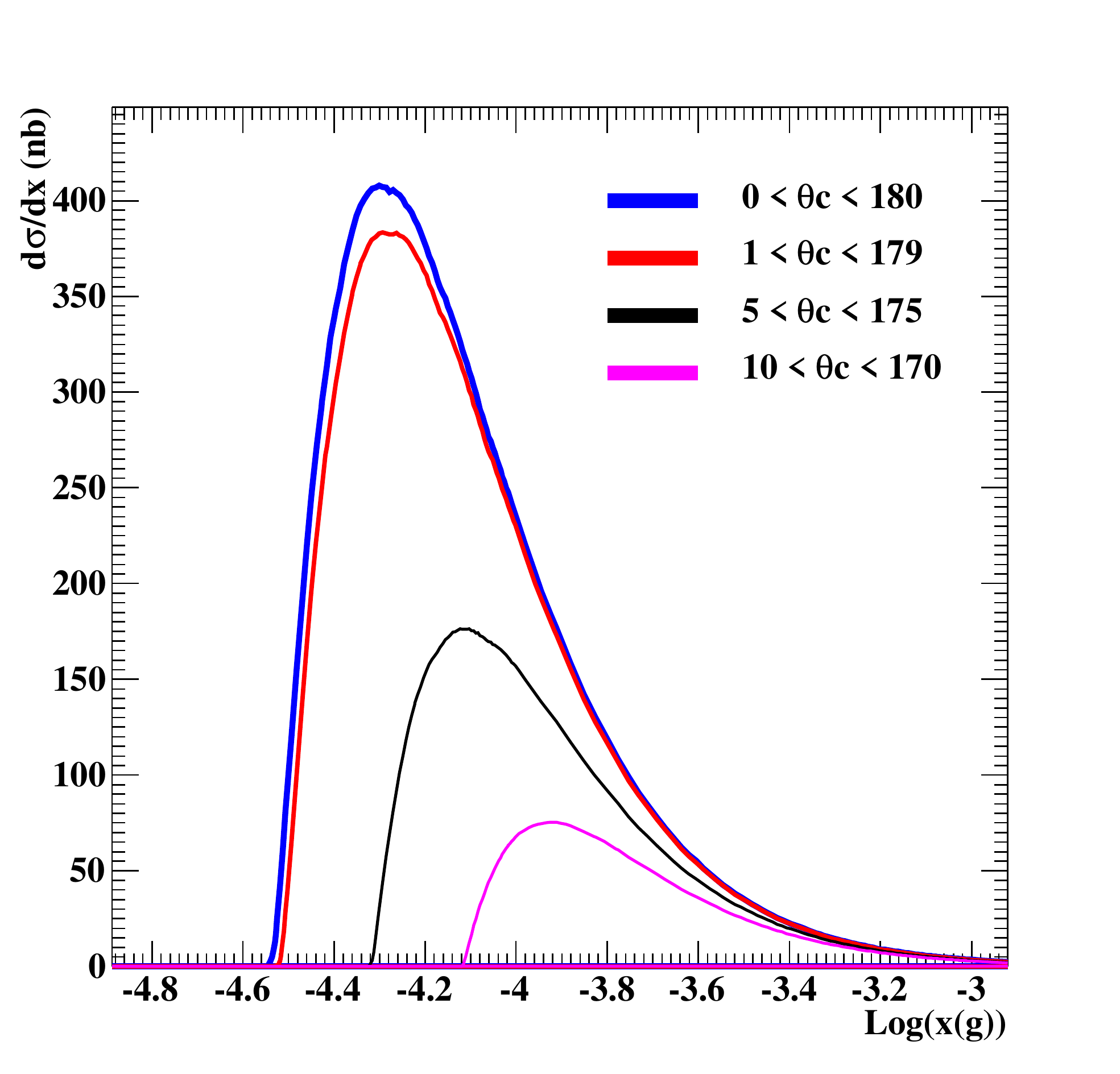}
\par\end{centering}

\caption{The effect of angular reach for $c\bar{c}$ (left) and $b\bar{b}$
(right) final states produced via CBS at the LHeC-2.}

\label{FIGURE7}
\end{figure}

\begin{figure}[H]
\begin{centering}
\includegraphics[width=0.4\paperwidth]{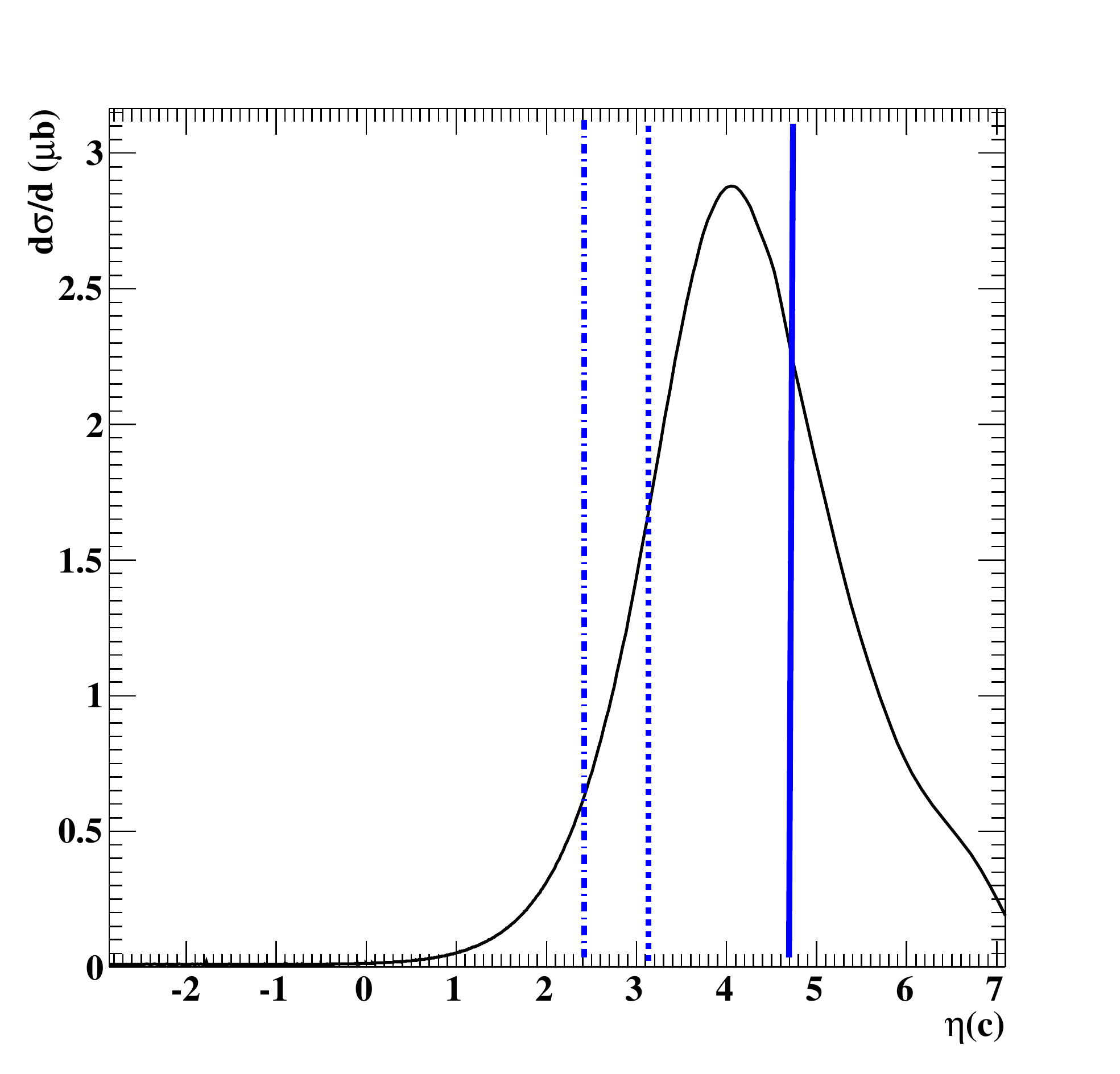}\includegraphics[width=0.4\paperwidth]{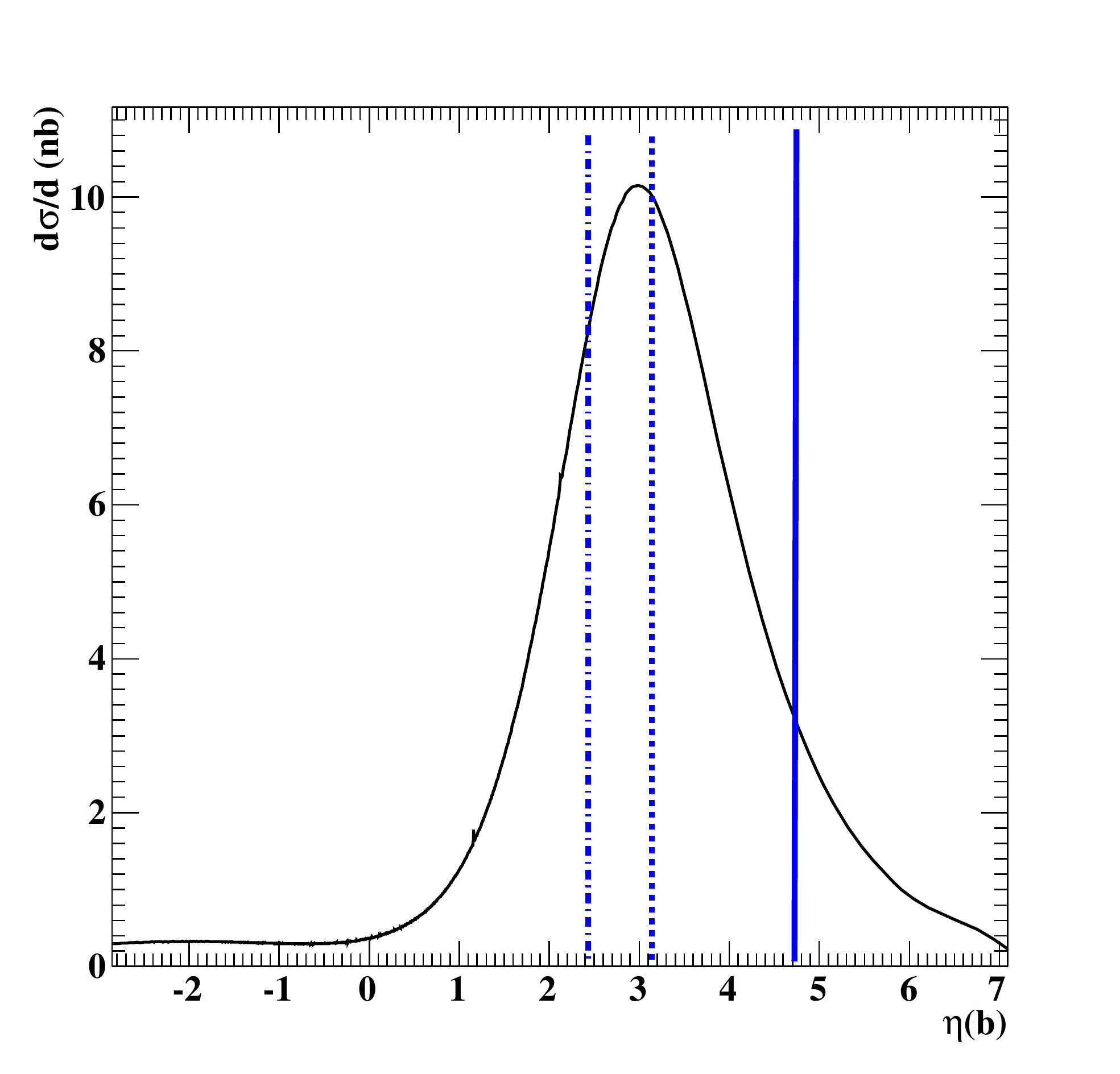} 
\par\end{centering}

\caption{The $\eta$ dependency of the $c\bar{c}$ (left) and $b\bar{b}$ (right)
production cross section via CBS at the LHeC-1. Vertical lines represent
$1{}^{o}$ (solid line), $5{}^{o}$ (dashed line) and $10{}^{o}$
(dot-dashed line) detector cuts.}

\label{Fig: eta_dependency_gp} 
\end{figure}

\begin{figure}[H]
\begin{centering}
\includegraphics[width=0.4\paperwidth]{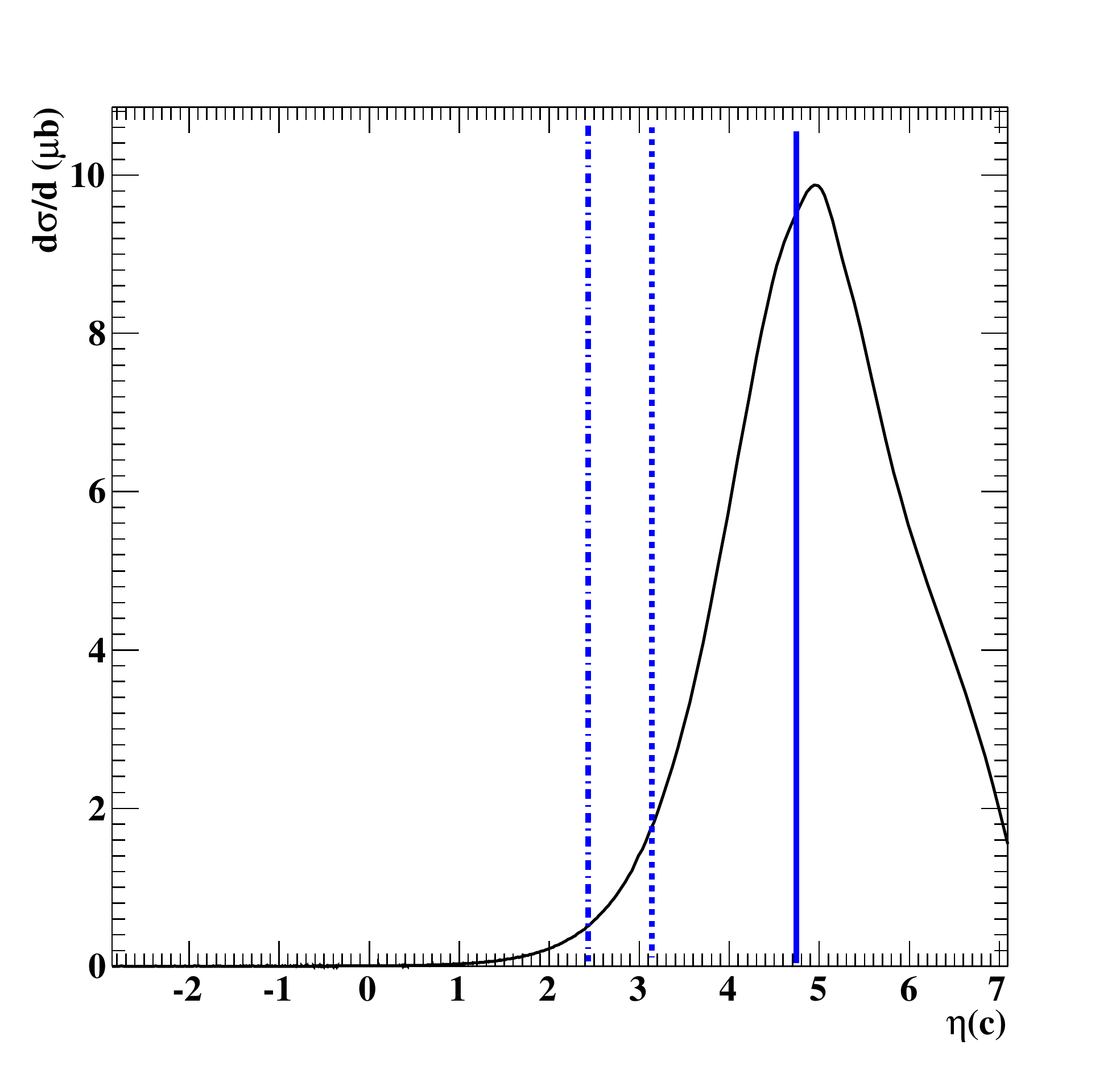}\includegraphics[width=0.4\paperwidth]{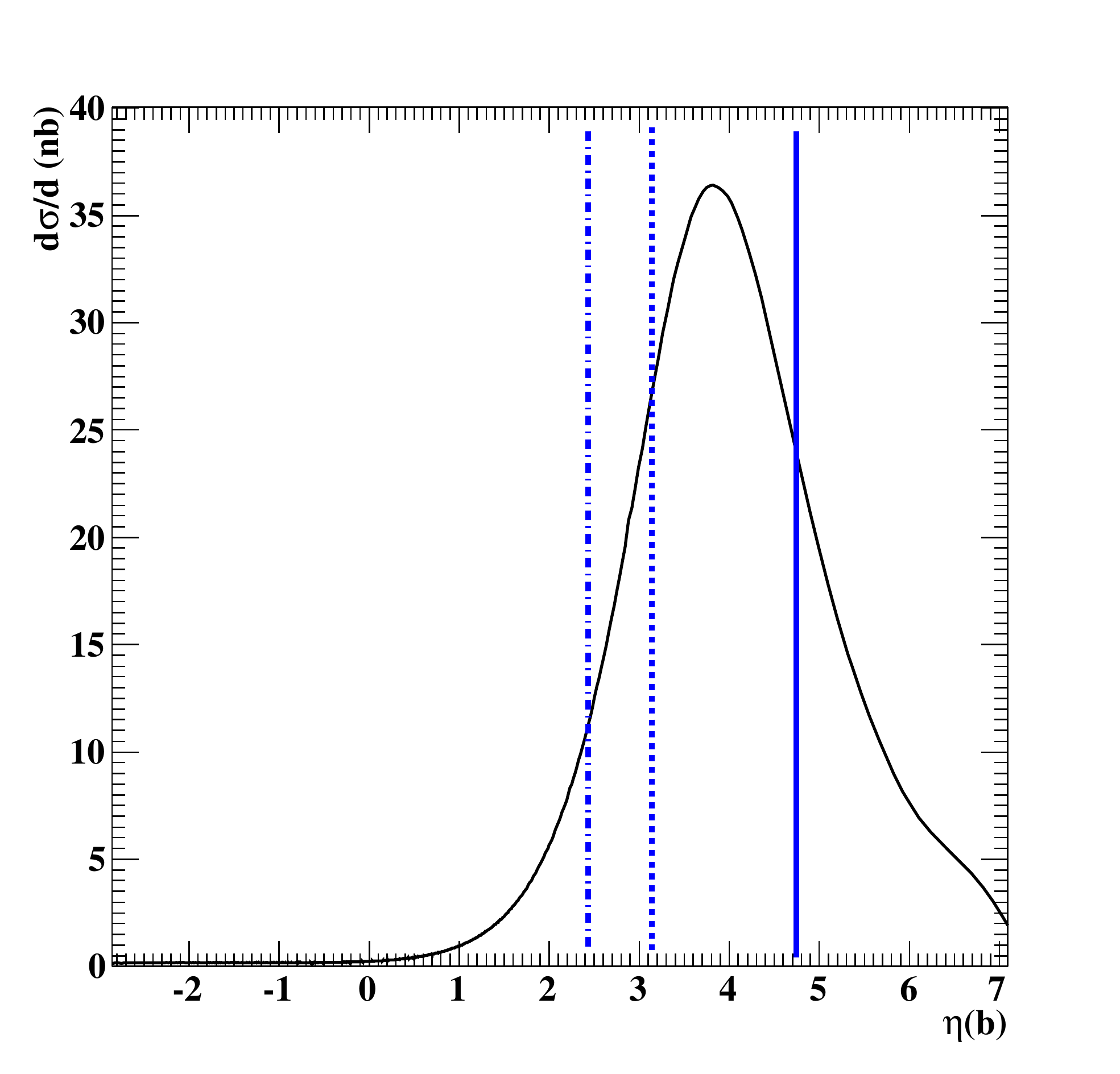}
\par\end{centering}

\caption{The $\eta$ dependency of the $c\bar{c}$ (left) and $b\bar{b}$ (right)
production cross section via CBS at the LHeC-2. Vertical lines are
same as in Fig. 8.}

\label{FIGURE9}
\end{figure}

The angular dependency of the relevant processes is important to estimate
the necessary $\eta$ coverage of the detector to be built and also
to estimate the eventual electron machine selection. For illustration
we consider $d\sigma/d\theta$ distribution where $\theta$ is the
angle between $c$ $(b)$ quark and proton beam direction. These distributions
for CBS at the LHeC-1 and LHeC-2 are presented in Figures \ref{FIGURE6}
and \ref{FIGURE7}, respectively. In Table \ref{TABLE 5}, we present
reachable $x(g)$ for different $\theta$ coverage. One can notice
that even for an angular loss of about 5 degrees, there is considerable
drop in both the cross section and in the $x(g)$ reach. This effect
can be understood by considering the $\eta$ dependence of the heavy
quark pair production cross section in $\gamma p$ collisions which
are shown in Figure \ref{Fig: eta_dependency_gp} and \ref{FIGURE9}.
The vertical solid line is representative for a 1 degree, the dashed
line for a 5 degree and the dot-dashed line is for 10 degree detector.
Therefore, in order to have a good experimental reach the tracking
should have an $\eta$ coverage up to 5. 

\begin{table}[H]
\caption{Reachable $x(g)$ for different $\theta$ coverage. }

\begin{centering}
\begin{tabular}{|c|c|c|c|c|}
\hline 
 & \multicolumn{2}{c|}{LHeC-1} & \multicolumn{2}{c|}{LHeC-2}\tabularnewline
\hline 
\hline 
 &  &  &  & \tabularnewline
$\theta$ & $c\bar{c}$ & $b\bar{b}$  & $c\bar{c}$ & $b\bar{b}$ \tabularnewline
\hline 
$0-180$ & $7.94\times10^{-6}$ & $6.91\times10^{-5}$ & $3.16\times10^{-6}$ & $3.02\times10^{-5}$\tabularnewline
\hline 
$1-179$ & $8.31\times10^{-6}$ & $6.91\times10^{-5}$ & $3.36\times10^{-6}$ & $4.36\times10^{-5}$\tabularnewline
\hline 
$5-175$ & $1.44\times10^{-5}$ & $7.94\times10^{-5}$ & $1.20\times10^{-5}$ & $4.78\times10^{-5}$\tabularnewline
\hline 
$10-170$ & $2.39\times10^{-5}$ & $1.00\times10^{-4}$ & $2.28\times10^{-5}$ & $7.58\times10^{-5}$\tabularnewline
\hline 
\end{tabular}
\par\end{centering}

\label{TABLE 5}
\end{table}

\section{Conclusions}

Measurements of $x(g)$ down to $3\times10^{-6}$ seem to be reachable
in $\gamma p$ collisions which is two order smaller than the HERA
coverage. These collisions provide higher cross section and better
$x(g)$ reach with respect to the $ep$ collisions with the same electron
beam energy. For the low $x(g)$ region, the enhancement factor compared
to the DIS $ep$ collisions is about 700 for $c\bar{c}$ final states
and about 230 for $b\bar{b}$ final states at the LHeC-2. Therefore,
for the final states with heavy quarks, even if the $\gamma p$ luminosity
is 10 times smaller than $ep$ luminosity (ERL option), the expected
number of events in $\gamma p$ collisions would be 70 and 20 times
higher than in $ep$ collisions for $c\bar{c}$ and $b\bar{b}$ final
state respectively. The enhancement factor compared to WWA $ep$ collisions
is about 24 for both final states. The angular sensitivity is very
important for smallest $x(g)$ reach for either $e$ or $\gamma$
beams, therefore a detector with a pseudorapidity coverage up to $\eta=5$
is required. This coverage is already achieved at the ATLAS and CMS
experiments using forward detector components.

Finally, $ep$ option of LHeC will give an opportunity to shed light
on the small $x(g)$ dynamics which is crucial for clarifying the
QCD basics. On the other hand, the $\gamma p$ option of LHeC will
essentially enlarge the LHeC capacity on the subject. Therefore, one
pulse linac should be considered as a baseline for LHeC design. In
this case, a higher center of mass energies can be achieved by lengthening
of the electron linac which will provide an opportunity to investigate
smaller $x(g)$ region. The luminosity loss can be compensated using
energy recovery linac without re-circulating arcs \cite{Litvinenko}
which may provide luminosity values exceeding $L=$$10^{34}$$cm^{-2}$$s{}^{-1}$
even with a multi-hundred GeV electron linac.


\begin{thebibliography}{10}
\bibitem{ATLAS Higgs} ATLAS collaboration, G. Aad et al., \textit{Combined
search for the standard model Higgs boson using up to 4.9 $fb{}^{-1}$of
$pp$ collision data at $\sqrt{s}=7\; TeV$ with ATLAS detector at
the LHC, Phys. Lett.} \textbf{B 710} (2012) 49 {[}arxiv:1202.1408{]}.

\bibitem{CMS Higgs} CMS collaboration, S. Chatrchyan et al., \textit{Combined
results of searches for the standard model Higgs boson in $pp$ collisions
at $\sqrt{s}=7\; TeV$, Phys. Lett.} \textbf{B 710} (2012) 26 {[}arxiv:1202.1488{]}.

\bibitem{QCD 2004} S. Sultansoy, \textit{Linac-ring type colliders:
Second way to TeV scale}, Eur Phys. J C \textbf{33}, s01, s1064-s1066
(2004).

\bibitem{J. Phys. G: Nucl. Part. Phys. 39 (2012) 075001} J.L.Abelleira
Fernandez , C.Adolphsen , A.N.Akay et al., \textit{A Large Hadron
Electron Collider at CERN: Report on the Physics and Design Concepts
for Machine and Detector}, J. Phys. G: Nucl. Part. Phys. 39 (2012)
075001.

\bibitem{arXiv:1211.4831v1 [hep-ex] 20 Nov 2012} J.L.Abelleira Fernandez
, C.Adolphsen , P.Adzic et al., \textit{A Large Hadron Electron Collider
at CERN}, arXiv:1211.4831v1 {[}hep-ex{]} 20 Nov 2012.

\bibitem{CompHep} E. Boos et al. (CompHEP Collaboration), \textit{Automatic
computations from Lagrangians to events}, Nucl. Instrum. Meth. A 534,
250 (2004). 

\bibitem{A. N. Akay} A. N. Akay, H. Karadeniz, and S.Sultansoy, \textit{Review
of Linac-Ring Type Collider Proposals}, Int. J. Mod. Phys. A 25, 4589
(2010). 

\bibitem{S. I. Alekhin} S. I. Alekhin et al., \textit{PHYSICS AT
gamma p COLLIDERS OF TeV ENERGIES}, Int. J. Mod. Phys. A 6, 21 (1991).

\bibitem{A. K. Ciftci et al.} A. K. Ciftci et al., \textit{Main parameters
of TeV energy }$\gamma-p$\textit{ colliders}, Nucl. Instrum. Methods
Phys. Res., Sect. A 365, 317 (1995).

\bibitem{TESLA * HERA based} A. K. Ciftci, S. Sultansoy, and O. Yavas,
\textit{TESLA {*} HERA based }$\gamma-p$\textit{ and }$\gamma-A$\textit{
colliders}, Nucl. Instrum. Methods Phys. Res.,Sect. A 472, 72 (2001).

\bibitem{Conversion efficiency} H. Aksakal et al., \textit{Conversion
efficiency and luminosity for }$\gamma-p$\textit{ colliders based
on the LHC-CLIC or LHC-ILC QCD explorer scheme}, Nucl. Instrum. Methods
Phys. Res., Sect. A 576, 287 (2007).

\bibitem{I. F. Ginzburg} I. F. Ginzburg et al., \textit{Colliding
gamma e and gamma gamma Beams Based on the Single Pass e+ e- Accelerators.
2. Polarization Effects. Monochromatization Improvement}, Nucl. Instrum.
Methods Phys. Res., Sect. A 219, 5 (1984).

\bibitem{Principles of photon colliders} V. I. Telnov, \textit{Principles
of photon colliders}, Nucl. Instrum. Methods Phys. Res., Sect. A 355,
3 (1995).

\bibitem{CTEQ} J. Pumplin et al., \textit{New generation of parton
distributions with uncertainties from global QCD analysis}, JHEP 0207
(2002) 012.

\bibitem{Litvinenko} V. Litvinenko, \textit{LHeC with \textasciitilde{}100\%
energy recovery linac}, 2nd CERN-ECFA-NuPECC workshop on LHeC, Divonne-les-Bains
1-3 Sep (2009).\end{thebibliography}
\end{document}